\documentclass[
    reprint,
    nofootinbib,
    aps,
    prd,
    amsmath,
    floatfix,
    ]{revtex4-2}
\usepackage{bm}
\usepackage{graphicx}
\usepackage{hyperref}

\newcommand{\loss}[1]{\ensuremath{L^{(\mathrm{#1})}}}

\begin{document}

\title{Meta-learning and data augmentation for mass-generalised jet taggers}
\author{Matthew J.\ Dolan}
\email{matthew.dolan@unimelb.edu.au}
\affiliation{ARC Centre of Excellence for Dark Matter Particle Physics, School of Physics, The University of Melbourne, Victoria 3010, Australia}
\author{Ayodele Ore}
\email{ayodeleo@student.unimelb.edu.au}
\affiliation{ARC Centre of Excellence for Dark Matter Particle Physics, School of Physics, The University of Melbourne, Victoria 3010, Australia}

\date{\today}
\begin{abstract}
    Deep neural networks trained for jet tagging are typically specific to a narrow
    range of transverse momenta or jet masses. Given the large phase space that the
    LHC is able to probe, the potential benefit of classifiers that are effective
    over a wide range of masses or transverse momenta is significant. In this work
    we benchmark the performance of a number of methods for achieving accurate
    classification at masses distant from those used in training, with a focus on
    algorithms that leverage meta-learning. We study the discrimination of jets
    from boosted $Z'$ bosons against a QCD background. We find that a simple data
    augmentation strategy that standardises the angular scale of jets with
    different masses is sufficient to produce strong generalisation. The
    meta-learning algorithms provide only a small improvement in generalisation
    when combined with this augmentation. We also comment on the relationship
    between mass generalisation and mass decorrelation, demonstrating that those
    models which generalise better than the baseline also sculpt the background to
    a smaller degree. \end{abstract}

\maketitle

\section{Introduction}

The introduction of deep learning techniques into collider physics in the past
decade has had a significant impact upon the field. One of the main areas of
application has been in the identification, or tagging, of hadronic jets at the
Large Hadron Collider (LHC). Deep learning methods have been applied to heavy
flavour tagging, $W$ boson and top quark tagging, quark-gluon discrimination,
and the identification of the decays of new heavy resonances. Recent reviews
include\,\cite{Larkoski:2017jix, Guest:2018yhq, Albertsson:2018maf,
    Radovic:2018dip, Carleo:2019ptp, Bourilkov:2019yoi, Schwartz:2021ftp}, and
there are now numerous examples of experimental analyses using these
methods\,\cite{CMS:2020zge, ATLAS-dijet, ATLAS:2020pcy, ATLAS:2019vwv}.

Much of this work on jet tagging studies the performance of the network within
a small domain, often a relatively narrow window of jet transverse momentum,
$p_T$, or jet mass, $m_J$. This means that a network trained around a specific
value of $p_T$ or $m_J$ will only be highly performant in the region around
those values. This specificity of the network to the statistics of the training
sample is a classic limitation of neural
networks\,\cite{NIPS2006_a74c3bae,5995347}. Given the broad range of transverse
momenta and invariant masses which the LHC is capable of probing, this creates
a problem for the use of machine-learning based taggers for the entirety of the
available phase space.

If training data is available for the entire range over which a model will be
tested, this issue can be addressed in a number of ways. A collection of neural
networks can be trained, one for each window, but this requires large
computational overheads and leads to discontinuities in the tagger response as
domain boundaries are crossed. Alternatively, there exist domain
\emph{adaptation} strategies that can produce individual classifiers with
near-optimal performance in the neighbourhood of the training domain. Examples
include Refs.\,\cite{parameterized,Shimmin:2017mfk,MUST, pivot,
    Aguilar-Saavedra:2021utu} as well as mass-decorrelation methods\,\cite{uboost,
    boost-uniform,Dolen:2016kst,Moult:2017okx,Heimel:2018mkt,Englert:2018cfo,
    mass-agnostic, moment-decomposition, disco}. A comprehensive recent study of
the performance of both data and training augmentation techniques and their
implications for decorrelation is Ref.\,\cite{mass-agnostic}.	 However, these
approaches do not directly optimise for \emph{generalisation} -- that is to
say, for effective tagging at $p_T$ or $m_J$ values distant from the training
domain. Generalisation of this sort is useful for situations in which it is not
easy to obtain sufficient training data at every desirable evaluation domain.
An important example is weak or unsupervised learning on experimental data,
where the  data for a given process depends on the available integrated
luminosity and the cross-section of the process of interest. Even in
fully-supervised cases that make use of simulation, one may encounter
efficiency limitations due to restrictive phase space cuts or matching
requirements.

Our goal in this work is to benchmark the generalisation performance of a
number of different methods. As examples of data augmentation we use
planing\,\cite{deOliveira:2015xxd,Chang:2017kvc} and
zooming\,\cite{shower-uncertainties}, and for regularisation we use $L_1$
weight decay. We also study domain generalisation algorithms that leverage
meta-learning as in Refs.\,\cite{metareg, feature-critic, mldg, masf, dadg,
    metavib}. In contrast to the previous approaches, such algorithms directly aim
to produce a model whose learning generalises to domains that are \emph{unseen}
during training. Specifically, we present results for the
MetaReg\,\cite{metareg} and Feature-Critic\,\cite{feature-critic}
algorithms.\footnote{We have also implemented MLDG\,\cite{mldg}, but did not
    find it competitive with our baseline networks and so do not present results
    for it.}

Meta-learning is a machine learning paradigm in which part of a model's
training algorithm is itself optimised, with the goal of instilling some
desired behaviour in the model across a number of related
tasks\,\cite{10.5555/296635, MAML}. For a review and survey see
Ref.\,\cite{meta-survey}. Meta-learning can be naturally applied to a domain
generalization context, where the training of a model in one domain and its
evaluation in another becomes a data instance for the meta-optimisation. A
successful application of meta-learned domain generalisation to jet tagging
would allow future searches similar to Refs.\,\cite{ATLAS-dijet,
    Chatrchyan:2012tx, Aaltonen:2012qt} to train a single network instead of
multiple independent networks.

We apply the different methods to the task of boosted resonance tagging across
a large range of masses, focusing on the scenario where one trains a network on
data at low masses and evaluates it at higher masses. This is motivated by the
abundance of data at low invariant masses, while interesting new physics is
likely to appear at larger masses, albeit with a small cross-section. We seek
to discriminate a  massive $Z'$ vector boson from a background of QCD jets. We
assume the $Z'$ decays into light quarks, and that it is boosted so the decay
products are reconstructed within a single jet. We study $Z'$ masses between
150 and 600~GeV, dividing this range into windows separated by 50~GeV. The
performance of the new algorithms is compared against a na\"ively-trained
baseline model. We use this resonance tagging task as an initial exploration of
the concept, and discuss other possible uses in the Conclusions.

Ultimately, we find that zooming is sufficient to produce near-optimal
generalisation and that the meta-learning algorithms provide only a minor
advantage when used in combination with zooming. If jets are not zoomed, the
meta-learning algorithms behave similarly to the baseline and $L_1$
regularisation is able to maintain accurate classification at new masses.  In
this respect our results resemble those of Ref.\,\cite{mass-agnostic} in the
context of decorrelation. They found that data augmentation using planing and
another method based on principal component analysis worked as well as ML
methods based on boosted decision trees and adversarial networks.

This paper is organised as follows. In Section \ref{sec:dg}, we review methods
for generalisation in neural networks and introduce the meta-learning
strategies of Refs.\,\cite{metareg, feature-critic}. Sections \ref{sec:data}
and \ref{sec:model} outline the dataset generation and model implementations
respectively. The classification performance of the trained models is presented
in Section \ref{sec:results}. We analyse the correlations between network
predictions and the input jet mass in Section \ref{sec:correlations} before
providing concluding remarks and comments on outlook in Section
\ref{sec:conclusions}.

\section{Domain generalisation}
\label{sec:dg}

The goal of domain generalisation is to train a model using one or more
distinct domains such that its prediction accuracy generalises to \emph{unseen}
domains\,\cite{NIPS2011_b571ecea,dg-survey-1,dg-survey-2}. Implementing such an
algorithm requires a collection of datasets \begin{align}
    \mathcal{D}=\{D_i\}_{i=1}^{p}\ , \end{align}  where $D_i$ is a dataset (a set
of example/label pairs) containing data in domain $i$ and there are $p$ domains
available in total. In analogy to training and testing datasets, $\mathcal{D}$
is split into disjoint \emph{source} and \emph{target} subsets,
\begin{align}\label{eq:data-split} \begin{tabular}{lcr}
        $\mathcal{S}\cup\mathcal{T}=\mathcal{D}$ & and & $\mathcal{S}\cap\mathcal{T}=$
        \O \, .\end{tabular} \end{align}  Data from the source domains is used to train
the classifier, which is ultimately to be evaluated on data from the target
domains. We will primarily be interested in the case where $\mathcal{S}$
contains jets at low masses and $\mathcal{T}$ contains jets at high masses.

In order to avoid over-fitting to the source domains one must instill some
information about the relationship between domains in the training procedure.
Notable approaches include data augmentation, representation learning and
meta-learning. Adversarial architectures have been used in high-energy physics
but typically for mass decorrelation or domain
\emph{adaptation}\,\cite{pivot,adversarial-DA,mass-agnostic,Heimel:2018mkt}. On
the other hand Ref.\,\cite{dg-survey-2} claims that adversarial approaches
excel in domain adaptation but not necessarily domain generalisation. In this
work, we benchmark the generalisation performance of five methods: mass
planing\,\cite{deOliveira:2015xxd,Chang:2017kvc},
zooming\,\cite{shower-uncertainties},
$L_1$-regularisation\,\cite{Krogh92asimple}, MetaReg\,\cite{metareg}, and
Feature-Critic\,\cite{feature-critic}. The first two of these involve data
augmentation, the third regularisation and the final two meta-learning. We go
through them in turn.

\subsection{Mass planing}
Mass planing was one of the first methods for mass-decorrelation that was
explored in ML studies of jet physics. A dataset can be mass-planed by
assigning each jet in the training set a weight $w$ such that \begin{equation}
    w^{-1} = \frac{d\sigma_C}{dm_J}(m_J=m), \end{equation}

\begin{figure*}[tp!]
    \centering
    \begin{tabular}{c}
        \includegraphics[width=0.87\textwidth]{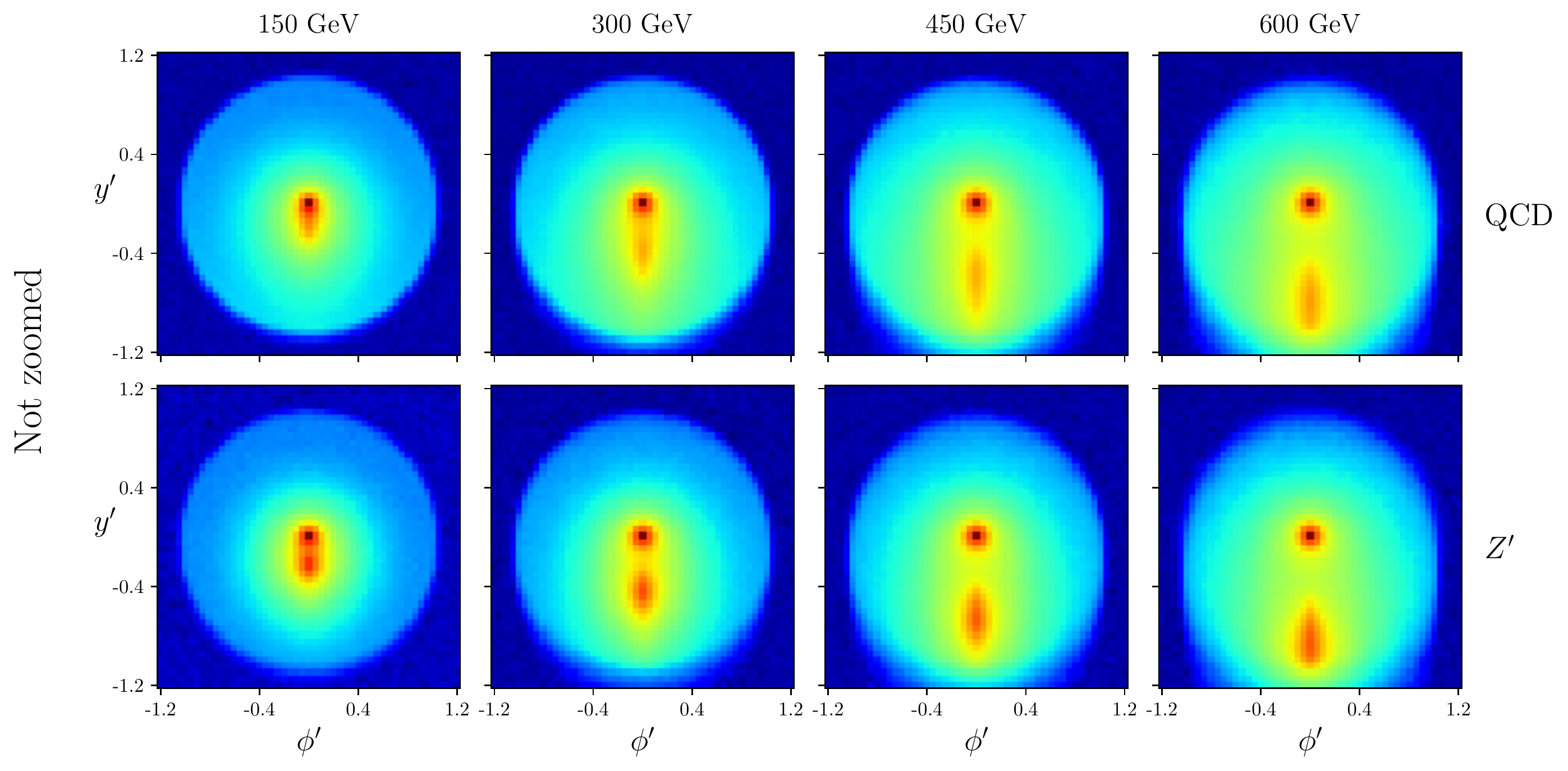}
        \\
        \includegraphics[width=0.87\textwidth]{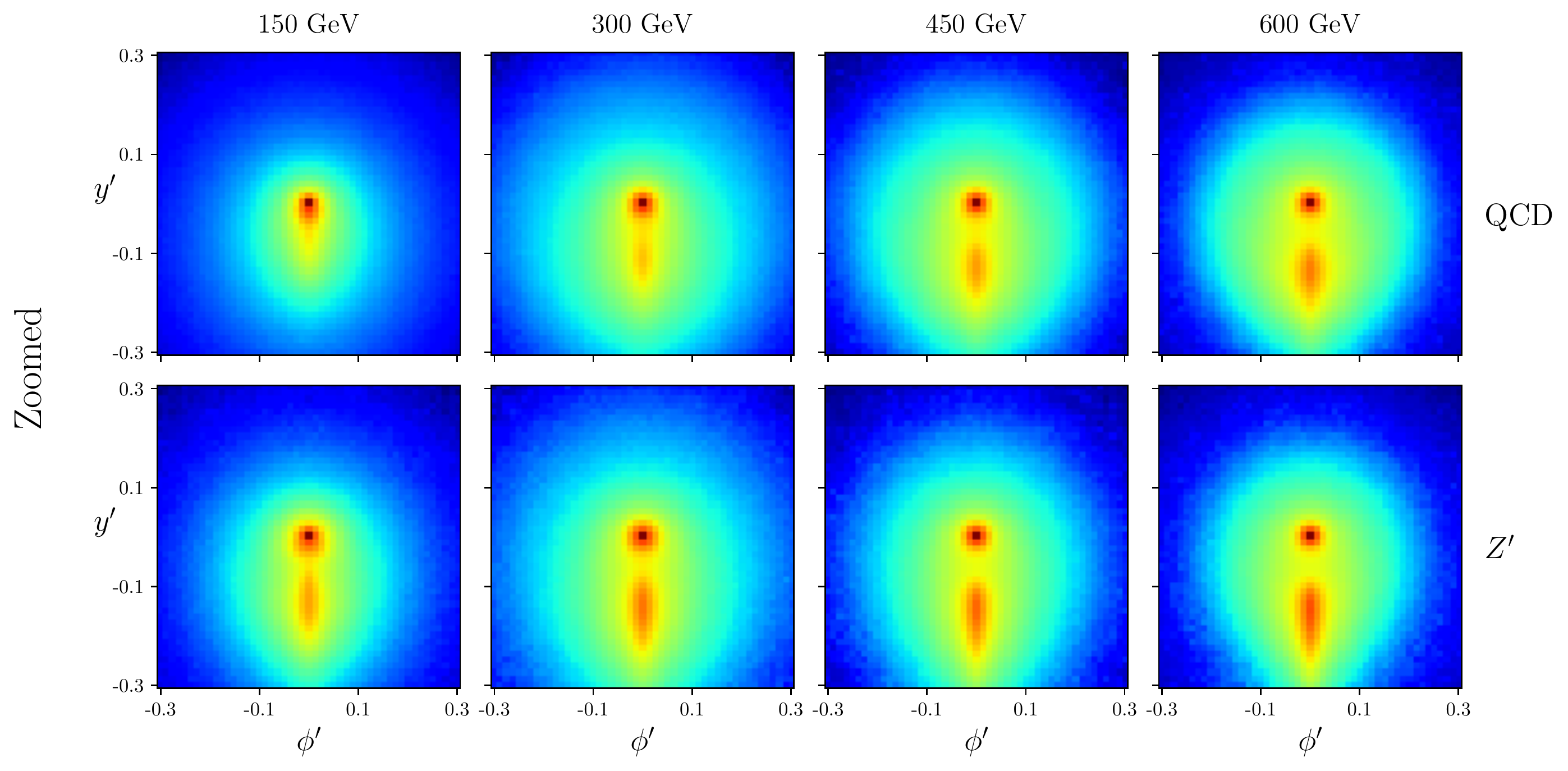}
    \end{tabular}
    \caption{Average jet images of the full datasets (500K events each)
        preprocessed according to Section \ref{sec:data} without (top) and with
        (bottom) the inclusion of zooming. The background is in the first and third
        lines, and the signal in the second and fourth lines. Zooming standardises the
        signal images over as the $Z'$ mass is varied (compare rows 2 and 4).}
    \label{fig:jet_images} \end{figure*}

where $C\in\{S,B\}$ is the appropriate signal / background class label for the
jet and $m$ is its mass. In practice, this simply corresponds to inverting the
$m_J$ histogram value at the bin in which the jet lies. After such a weighting,
the signal and background have uniform distributions in $m_J$ and as such the
jet mass no longer provides discrimination.\footnote{It is actually only
    necessary that the signal and background distributions match one another, not
    that they be uniform.}

While mass-planing is not strictly a generalisation technique,
mass-decorrelation methods can in general be expected to improve performance at
new masses since networks trained under these methods are discouraged from
using the jet mass to make predictions. The DDT\,\cite{Dolen:2016kst} method
has been used for this purpose in Ref.\,\cite{CMS:2019qem}.

\subsection{Zooming}
\label{sec:zooming}

One way in which jets from boosted resonances with distinct masses differ is
the separation of their two subjets, $\Delta R\sim2m/p_T$. A neural network
trained on jets in a narrow mass window does not learn this scaling
relationship, leading to poor generalisation at new masses. The potential
benefit of scale-invariant jet tagging has been noted in
Refs.\,\cite{shower-uncertainties,Gouzevitch:2013qca,Dolen:2016kst}, motivating
the addition of a \emph{zooming} transformation to the jet preprocessing steps.

The implementation in Ref.\,\cite{shower-uncertainties} reclusters the jet into
subjets and then scales their separation by a factor dependent on the jet mass
and transverse momentum. For application to widely-varying jet masses, we use a
different procedure in which no reference to $m_J$, jet $p_T$ or clustering
radius is used. Specifically, we scale the $\eta$ and $\phi$ coordinates of all
jet constituents by a factor that ensures the average distance between each
particle and the leading constituent is 0.1. We find that this variant of
zooming is effective, demonstrated by the averaged jet images for our datasets
shown in Fig.\,\ref{fig:jet_images}. The top two rows show the background and
signal images without zooming, and the bottom two are the images after zooming
has been applied. Comparing the second and fourth rows, the zooming procedure
standardises the signal images over a range of $Z'$ masses.

\subsection{Regularisation}
Regularisation may be defined as any modification to a learning algorithm that
reduces generalisation error, even at the expense of increased training error.
In this class, the most common techniques used in modern neural networks
include weight decay ($L_1$ or $L_2$ regularisation\,\cite{Krogh92asimple}),
Batch Normalisation\,\cite{DBLP:journals/corr/IoffeS15},
Dropout\,\cite{Srivastava2014DropoutAS} and
DropConnect\,\cite{Wan2013RegularizationON}. While these methods lead to
improved generalisation error for testing samples drawn from the same
distribution as the training data, they are not necessarily effective at domain
generalisation.

In this work, we implement $L_1$ regularisation, where the classification loss
$\mathcal{L}$ is extended as \begin{equation}\label{eq:L1}
    \mathcal{L}\rightarrow\mathcal{L}+\lambda\sum_i\left|\theta_i\right|,
\end{equation} where $\theta_i$ are the neural network weights and $\lambda$ is
a hyperparameter that sets the regularisation strength. One can similarly
define $L_2$ regularisation by \begin{equation}\label{eq:L2}
    \mathcal{L}\rightarrow\mathcal{L}+\lambda\sum_i{\theta_i}^2. \end{equation}
Gradient updates derived from these losses include a term that reduces the size
of the weights $\theta$, thus mitigating specificity and, thereby, overfitting.
The difference between the $L_1$ and $L_2$ schemes is that the former allows
weights to be sent to zero while the contributions to the gradient from the
latter are proportional to the size of the weight, which damps the decay. We
have also implemented and studied the generalisation behaviour of $L_2$
regularisation. However, we find that for the task we study $L_2$
regularisation does not generalise as well as $L_1$, and so we only present
results for $L_1$ regularisation.

\subsection{Meta-learning}
\label{sec:meta-learning}

\begin{figure*}
    \centering
    \includegraphics[width=\textwidth]{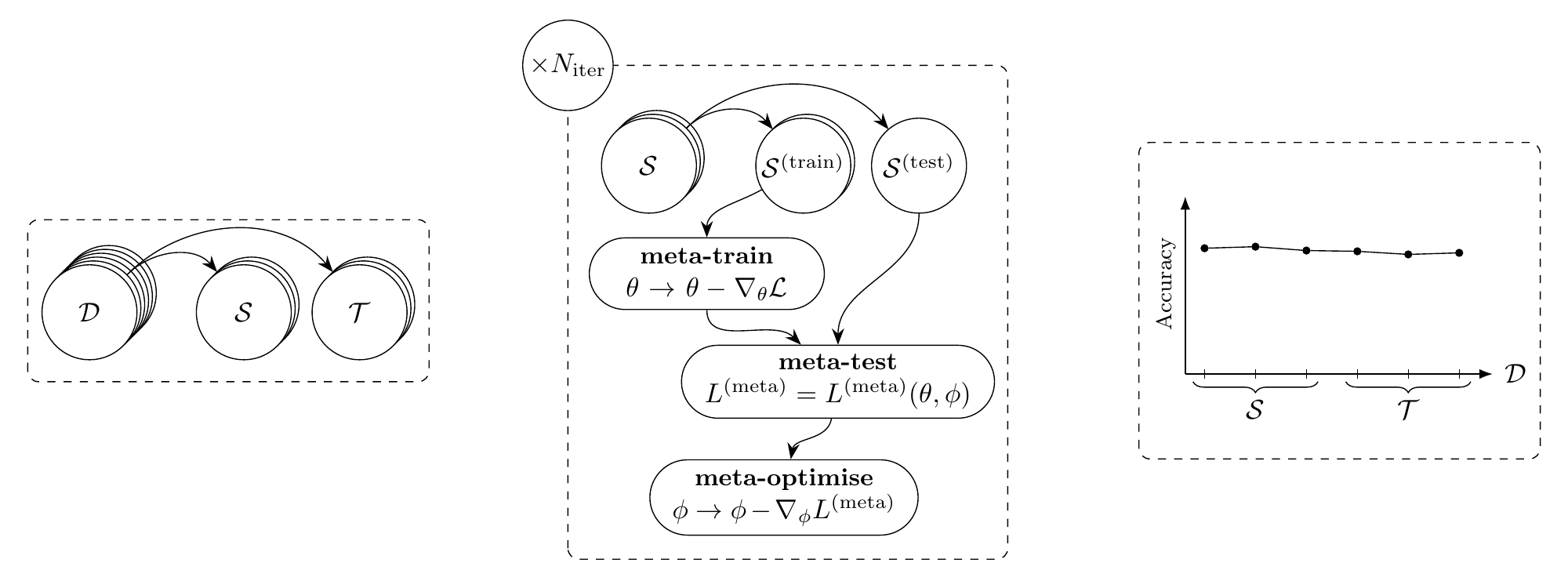}
    \caption{General schematic for domain generalisation via meta-learning. A
        collection of datasets $\mathcal{D}$ spread across the domain is split into
        source and target collections. The parameters $\phi$ of the meta-representation
        are updated using data from the source domains only. A particular algorithm
        conducts meta-training, meta-testing, and meta-optimisation at each iteration.
        Finally, the model is evaluated on the entire collection $\mathcal{D}$ and the
        degree of generalisation achieved can be measured.} \label{fig:meta-diagram}
\end{figure*}

Meta-learning is a machine learning paradigm in which a model's training
algorithm is itself optimised, typically to achieve some goal over a number of
related tasks. For domain generalisation, the tasks differ only by shifts in
the given domain and the goal is to improve performance on unseen domains.

The general prescription for meta-learning is as follows. One first
parametrises a component of the base model's training algorithm by some
variables~$\phi$. This component is referred to as the
\emph{meta-representation} and can take a variety of forms such as the initial
network weights\,\cite{MAML, mldg}, the loss function\,\cite{feature-critic},
the optimiser\,\cite{DBLP:journals/corr/AndrychowiczDGH16}, a
regulariser\,\cite{metareg} or even the model architecture itself. For a more
exhaustive categorisation, see Ref.\,\cite{meta-survey}. The parameters of the
meta-representation are updated iteratively, with each step divided into three
stages: \emph{meta-training}, \emph{meta-testing} and \emph{meta-optimisation}.
During meta-training, the base classifier is updated via gradient descent using
the meta-representation variables $\phi$. The model is then evaluated in the
meta-test stage by calculating a \emph{meta-loss} \loss{meta} that encapsulates
the meta-learning objective. Since \loss{meta} depends on the parameters $\phi$
through the meta-train phase, back-propagation can be used to find a direction
$d\phi$ that gives improved performance with respect to the goal. The variables
$\phi$ are updated in this direction during meta-optimisation.

Fig.\,\ref{fig:meta-diagram} illustrates how meta-learning operates in a domain
generalisation context. In this case, meta-training and meta-testing are
conducted in disjoint subsets of the source domains, denoted
$\mathcal{S}^{(\mathrm{train})}$ and $\mathcal{S}^{(\mathrm{test})}$
respectively. In this way, source/target domain shift is simulated as the model
is trained. By taking the meta-loss to be the classification loss of the model
on $\mathcal{S}^{(\mathrm{test})}$, the meta-representation parameters $\phi$
are optimised to enforce robust performance in unseen domains. The
representation can then be deployed on a new model to be trained on all source
domains, and is expected to achieve improved performance on the target domains
compared with a model trained in a na\"ive fashion.

Below, we introduce the two meta-learning algorithms that we benchmark in this
work: MetaReg\,\cite{metareg} for which the meta-representation is a weight
regulariser and Feature-Critic\,\cite{feature-critic} for which the
meta-representation is an auxiliary loss function that depends on the
latent-space features produced by the model. We also implemented the MLDG
algorithm\,\cite{mldg}, which  meta-learns the network's weight initialisation,
but found no difference in performance compared to the baseline. Accordingly we
do not include results for MLDG in our plots.  Other examples of domain
generalisation via meta-learning that could also be applied include
Refs.\,\cite{masf,dadg,metavib}.

\subsubsection*{Meta-Regularisation}
The MetaReg\,\cite{metareg} algorithm uses a regulariser $R_\phi$ as the
meta-representation. In Ref.\,\cite{metareg} three parametrised regularisers
were used: \emph{weighted} $L_1$ and $L_2$ (in which the sums in Equations
\ref{eq:L1} and \ref{eq:L2} are weighted by the learned parameters $\phi$) and
a multi-layer perceptron (MLP).

To avoid memory limitations for large models (tracking higher-order gradients
through many updates is costly) the learned regulariser acts only on a subset
of the network layers. As such the base model is split into two subnetworks:
the \emph{feature extractor} $F_\psi$ and the \emph{task network} $T_\theta$.
The full network is the composition $T_\theta\circ F_\psi$ and regularization
is only applied to task network weights $\theta$. During optimisation of the
regulariser, a single feature extractor is updated from all source domains,
while independent task networks are used in each domain. In this way, only the
task networks learn domain-specific information and this is what allows the
regulariser to enforce domain invariance.

Training the meta-regulariser proceeds by randomly initialising weights $\phi$,
$\psi$ and $\theta_i$ for the regulariser, feature extractor and
$i$\textsuperscript{th} task network respectively, before performing the
following steps at each iteration. \begin{enumerate} \item In each source domain $D_i$, perform $k$ updates of the network $T_i\circ F$
          without regularisation: $$\left.\begin{matrix}\psi \rightarrow \psi -
                  \alpha\nabla_{\psi}\mathcal{L}^{(i)}_{\psi,\theta_i} \\[8pt]\theta_i
                  \rightarrow \theta_i -
                  \alpha\nabla_{\theta_i}\mathcal{L}^{(i)}_{\psi,\theta_i}\end{matrix}\ \right\}\times
              k\,.$$ where $\mathcal{L}^{(i)}_{\psi,\theta}=\mathcal{L}(y,
              T_{\theta_{i}}\circ F_\psi({\bf x}))$ is the classification loss of the model
          evaluated on ${\bf x},\,y \sim D_i$.

    \item Randomly select $a,b\in\{1,\:\cdots,p\}$ (where $p$ is the number of domains)
          such that $a\neq b$ and initialize a new task network $\Hat{T}$ with parameters
          $\theta_a$: $$\Hat{\theta}\rightarrow\theta_a\,.$$

    \item {\bf Meta-train} on $D_a$ by performing $l$ updates to $\Hat{\theta}$ for
          the network $\Hat{T}\circ F$ using the regularised loss:
          $$\left.\begin{matrix}
                  \Hat\theta \rightarrow \Hat\theta -
                  \alpha\nabla_{\Hat{\theta}}\left(\mathcal{L}^{(a)}_{\psi,\Hat\theta}+R_\phi(\Hat{\theta})\right)\end{matrix}\ \right\}\times l\,.$$

    \item {\bf Meta-test} by calculating the \emph{unregularised} loss on $D_b$
          using $\psi$ and $\Hat{\theta}$: $$\loss{meta} =
              \mathcal{L}^{(b)}_{\psi,\Hat{\theta}}$$

    \item {\bf Meta-optimise} by updating the regulariser parameters using the
          meta-test loss: $$\phi \rightarrow \phi - \beta\nabla_\phi \loss{meta}\,.$$
\end{enumerate} In the above, $\alpha$ and $\beta$ are learning rates. Once the
specified number of meta-optimisation steps has been completed, the regulariser
weights $\phi$ are frozen. A new model is then initialised using the trained
feature extractor and random task network. The model is then trained to
minimise the regularised loss using any optimisation algorithm.

\subsubsection*{Feature-Critic}
The Feature-Critic algorithm was introduced in Ref.\,\cite{feature-critic},
which addresses the heterogeneous domain generalisation problem wherein the
label spaces are not shared between domains. It is compatible, however, with
the more common homogeneous case that we explore here.

This algorithm learns a contribution to the loss function that aims to promote
cross-domain performance and, similarly to MetaReg, it considers the model
separated into a feature extractor, $F_\psi$ and a task network $T_\theta$. The
meta-representation is a \emph{feature critic} denoted $h_\phi$ which acts on
the features produced by $F$. The critic is trained alongside the feature
extractor and task network in the following way before its parameters are
frozen and the full model is fine-tuned.

\begin{enumerate}
    \item  Randomly split $\mathcal{S}$ into $\mathcal{S}^{(\mathrm{train})}$ and
          $\mathcal{S}^{(\mathrm{test})}$ in the same fashion as
          Eq.\,\ref{eq:data-split}, ensuring $\mathcal{S}^{(\mathrm{train})}$ contains
          $S$ domains. \item {\bf Meta-train} by calculating classification and auxiliary
          losses on $\mathcal{S}^{(\mathrm{train})}$ and defining updated weights: $${\bf
                      x},\,y \sim \mathcal{S}^{(\mathrm{train})}$$ $$\psi_{\mathrm{old}} = \psi -
              \gamma\nabla_\psi\mathcal{L}^{(\mathrm{train})}_{\psi,\theta}$$
          $$\psi_{\mathrm{new}} = \psi_{\mathrm{old}} - \gamma\nabla_\psi
              h_\phi(F_\psi({\bf x})) \,.$$ \item {\bf Meta-test} by sampling
          $\mathcal{S}^{(\mathrm{test})}$ and calculating the improvement in
          classification provided by the critic, $$\loss{meta} =
              \tanh\left(\mathcal{L}^{(\mathrm{test})}_{\psi_\mathrm{new},\theta} -
              \mathcal{L}^{(\mathrm{test})}_{\psi_\mathrm{old},\theta}\right) \,.$$

    \item  {\bf Meta-optimise} by updating the critic parameters. The feature
          extractor and task network are also updated here: $$\psi\rightarrow\psi -
              \alpha\nabla_\psi\left(\mathcal{L}^{(\mathrm{train})}_{\psi,\theta}+h_\phi(F_\psi({\bf
                      x}))\right)$$ $$\theta\rightarrow\theta -
              \alpha\nabla_\theta\left(\mathcal{L}^{(\mathrm{train})}_{\psi,\theta}+h_\phi(F_\psi({\bf
                      x}))\right)$$ $$\phi\rightarrow\phi - \beta\nabla_\phi\loss{meta} \,.$$
\end{enumerate} In the above algorithm, $\alpha$, $\beta$ and $\gamma$ are
learning rates.

\section{Simulation details}
\label{sec:data}

In this work, we train networks to discriminate signal jets produced by a
boosted hypothetical $Z'$ boson which decays into a pair of light quarks,
against a QCD jet background. We consider a range of potential resonance masses
$m_{Z'}$. We focus on  a relatively light (sub-TeV), boosted $Z'$ whose origin
may be in the decay of a heavier multi-TeV resonance. Accordingly, we are
interested in the situation where the decay products from the $Z'$ decay are
reconstructed within the same large radius jet.

To generate the datasets, signal and background processes are simulated with
    {\sc MadGraph5 2.8.0}\,\cite{madgraph}, showered in {\sc Pythia
        8.244}\,\cite{pythia} then passed to {\sc Delphes 3.4.2}\,\cite{delphes} with
the default CMS card. For the signal, we use a simple $Z'$ model from {\sc
        FeynRules}\,\cite{feynrules}. The signal and background jets are obtained from
$pp\to Z(\nu\Bar{\nu})Z'$ and $pp\to Z(\nu\Bar{\nu})j$ processes respectively.
We produce 10 datasets, corresponding to masses $m_{Z'}=150$ GeV to
$m_{Z'}=600$ GeV at $50$ GeV intervals, each of which contains $5\times 10^5$
$Z'$ and $5\times 10^5$ QCD events. Thus the full collection of datasets is
$\mathcal{D}=\{D_{150}, D_{200},\:\cdots,D_{600}\}$ where the labels indicate
the $Z'$ mass. The $Z'$ widths are calculated automatically by {\sc MadGraph},
which with the default couplings gives $\Gamma_{Z'}/m_{Z'}\sim 0.0275$ at every
mass.

Within {\sc Delphes}, events are clustered by {\sc FastJet 3.3.4} using the
anti-$k_t$ algorithm with $R=1.0$ and the leading jet is selected. Parton-level
cuts of $|\eta_J|, |\eta_{Z'}|<2.0$ and $\not\!\!E_T>1.2$~TeV are applied in
    {\sc MadGraph}, the latter of which boosts the jets to ensure that decay
products from the $Z'$ fall within a cone of appropriate radius for jet
reconstruction. This is motivated by simplicity and experimental searches,
which usually do not vary the jet radius. Since this missing energy cut is the
same for all datasets, jets in different datasets will	    have different
boosts, which may be considered a component of the domain shift which our
networks will learn to generalise. Cuts of $|m_J-m_{Z'}|<m_{Z'}/4$ and
$p_{T}>1.2$~TeV are applied after detector simulation. The size of the jet mass
window varies to account for the scaling of the ${Z'}$ width with its mass.
This causes overlaps between nearby domains which results in a jagged QCD jet
mass distribution for aggregated datasets. Although this can be resolved by an
appropriate reweighting of events that takes into account both the number of
mass windows in which a jet lies as well as the relative cross-section between
windows, such a reweighting increases domain specificity so we do not explore
it here.

Jets are preprocessed by first translating all constituents in the
rapidity-azimuth plane such that the leading constituent lies at the origin. A
rotation is then applied to position the centre of momentum below the origin.
Jets are then optionally zoomed as per Section\,\ref{sec:zooming}. Each jet is
stored as a list of constituent information in the format $(p_T/p_{T,J}, \eta',
    \phi', g(\text{\texttt{pdg\_id}}))$ where $\eta'$ and $\phi'$ respectively
denote pseudorapidity and azimuthal angle coordinates after the mentioned
transformations and $g$ maps Particle Data Group Monte Carlo numbers to small
floats beginning at $0.05$ and incrementing $0.1$ for each class of
constituent. The arrays are serialised and saved in {\tt TFRecord} format
allowing for efficient interfacing with {\sc  Tensorflow}\,\cite{tensorflow},
which is necessary to avoid memory issues associated with loading a large
number of datasets. In order to facilitate future studies on mass
generalisation, the complete collection of datasets has been made available at
Ref.\,\cite{mass-generalisation-dataset}.

\section{Model details}
\label{sec:model}
\begin{table*}
    \begin{ruledtabular}
        \begin{tabular}{ccc}
            {\bf Hyperparameter} & {\bf MetaReg}     & {\bf Feature-Critic} \\\hline
            Iterations           & $10^{\,3}$, $\bm{5\times10^{\,3}}$, $10^{\,4}$,  $5\times10^{\,4}$ & $10^{\,3}$, $5\times10^{\,3}$, $\bm{10^{\,4}}$,  $5\times10^{\,4}$    \\ $\beta$    &
            $10^{ -1}$, $10^{ -2}$, $\bm{10^{ -3}}$, $10^{ -4}$, $10^{ -5}$, $10^{ -6}$           & $10^{ -1}$, $10^{ -2}$, $\bm{10^{ -3}}$, $10^{ -4}$, $10^{ -5}$, $10^{ -6}$                                \\ Regulariser & {\bf Weighted} $\bm{L_1}$, Weighted $L_2$, MLP
                                 & -                                        \\ $k$ & 1, $\bm{4}$, 16, 32 & -
            \\ $l$ & $\bm{1}$, 4, 8, 16 & -					  \\
            $\gamma$             & -                 & $10^{ -1}$, $10^{ -2}$, $10^{ -3}$, $10^{ -4}$, $\bm{10^{ -5}}$, $10^{ -6}$
            \\ $S$ & - & $\bm{1}$, 2 \\\end{tabular} \end{ruledtabular} \caption{A summary of the hyperparameters for MetaReg and Feature-Critic that were explored by grid search. Selected parameters are shown in bold. The learning rate
        hyperparameter $\alpha$ is fixed to $10^{-3}$ to match the baseline.}
    \label{tab:hyperparameters}
\end{table*}

\begin{figure*}
    \centering
    \includegraphics[width=\textwidth]{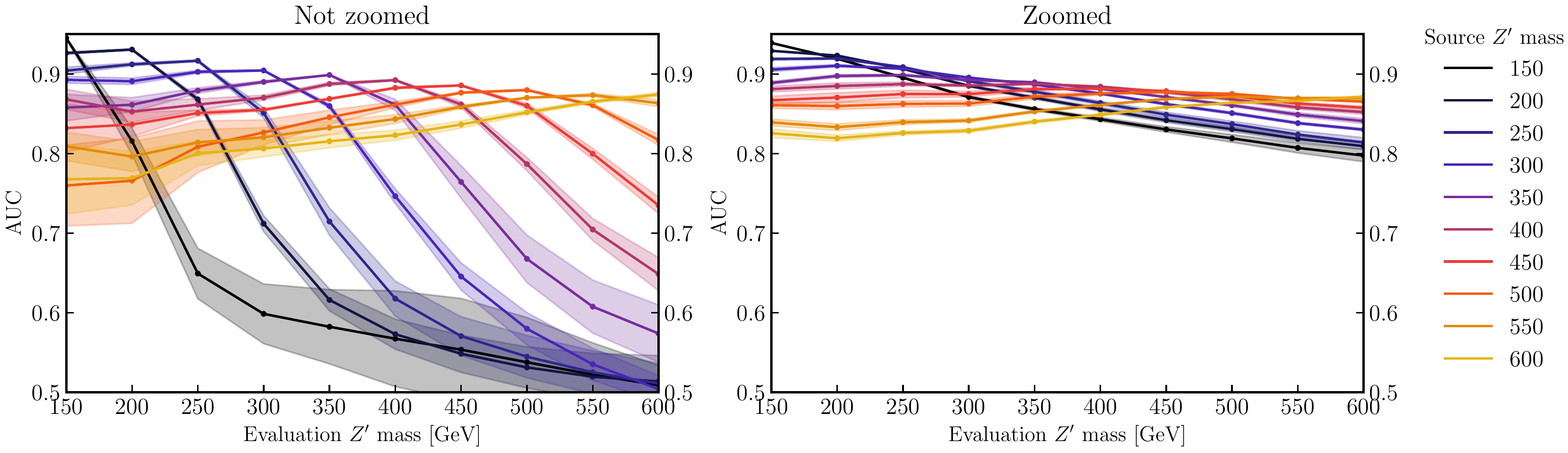}
    \caption{The performance of baseline PFNs trained on single datasets with
        (right) and without (left) zooming. Zooming improves generalisation for all
        source masses, particularly for networks trained at low masses which generalise
        to high masses. Curves are the mean of 5 independent networks with bands
        displaying one standard deviation around the mean.} \label{fig:baseline-single}
\end{figure*}

We will compare the performance of the approaches from Section \ref{sec:dg}
using a Particle Flow Network (PFN)\,\cite{pfn} as the underlying classifier.
PFNs are an implementation of the DeepSets\,\cite{deep-sets} framework and
treat jets as point clouds, where each point is a jet constituent. Each
particle in the jet is mapped into a latent space with a per-particle network.
The outputs of these networks are then summed over to obtain a latent
representation of the whole jet, which is then mapped by another jet-level
network to an output giving the value of the learned observable. For all
algorithms we study, the PFN has a per-particle network with layer sizes
(100,100,256) and a jet-level network with layer sizes (100,100,100,2). All
parameters use Glorot uniform initialisation\,\cite{Glorot10} and all
activations are rectified linear units (ReLU) except for the two-unit output
layer for which we which we use a softmax function.

For training the PFNs, we use the cross-entropy classification loss optimised
via AMSGrad\,\cite{amsgrad} with learning rate $\alpha=10^{-3}$ and a
mini-batch size of 200.\footnote{Despite the results of studies such as
    Refs.\,\cite{DBLP:journals/corr/abs-1712-07628, Wilson2017TheMV} which note
    that adaptive gradient methods achieve poorer generalisation, we did not
    observe any difference in results when using SGD.} For $L_1$-regularised
training, we use a weight of $10^{-3}$ and the penalty applies only to the
jet-level network. Weights for mass planing are calculated from a 64-bin
histogram and normalised to have unit mean per batch. With the exception of the
meta-learning algorithms, which consume data from different source domains as
outlined in Section\,\ref{sec:meta-learning}, the networks are fit on all
source domains as a single aggregated dataset. We also trained PFNs on each
domain individually with the same settings. In all cases, we use a
training/validation/testing split of 0.75/0.1/0.15 and the PFN is trained for a
maximum of 100 epochs. If the validation loss has not decreased for 5 epochs,
training is halted and the best weights are restored.

We implement the MetaReg and Feature-Critic algorithms in {\sc TensorFlow} and
fix their $\alpha$ learning rates to $10^{-3}$ to match the baseline PFN. In
both algorithms, the feature extractor $F_\psi$ is taken to be the per-particle
network (including the latent-space sum) and the task network $T_\theta$ is the
jet-level network. For Feature-Critic, the auxiliary loss $h_\phi$ has two
hidden layers with 512 and 128 nodes with ReLU activations. The scalar outputs
of $h_\phi$ and the MetaReg regulariser $R_\phi$ are passed through a softplus
function to ensure a convex loss. While no such activation is applied to
$R_\phi$ in Ref.\,\cite{metareg}, without it the loss function is unbounded
from below resulting in unstable training. Meta-optimisation is performed using
AMSGrad and the remaining hyperparameters are selected via a grid search on
$\mathcal{S}=\{D_{150},D_{200},D_{250}\}$ with zooming which we summarise in
Table\,\ref{tab:hyperparameters}. We adopt those parameters that produce the
greatest AUC score on $D_{300}$.

\section{Results and Discussion}
\label{sec:results}

In this Section, we present the performance of the trained models under various
generalisation settings. Firstly, we demonstrate the benefit of zooming jets in
Fig.\,\ref{fig:baseline-single}, which shows the AUC scores of PFNs trained
with baseline settings on an individual source domain without (left panel) and
with zooming (right panel). In the left panel we see that the performance of
each network degrades substantially away from the domain in which it was
trained. A network trained in the $m_{Z'}=150$~GeV domain is only slightly
better than a random classifier at 600~GeV.  The maximum AUC achieved at each
mass decreases at larger masses due to the fact that high mass QCD jets have
stronger pronged structure and thus the discrimination task is inherently more
difficult in these datasets.

\begin{figure*}
    \centering
    \includegraphics[width=\textwidth]{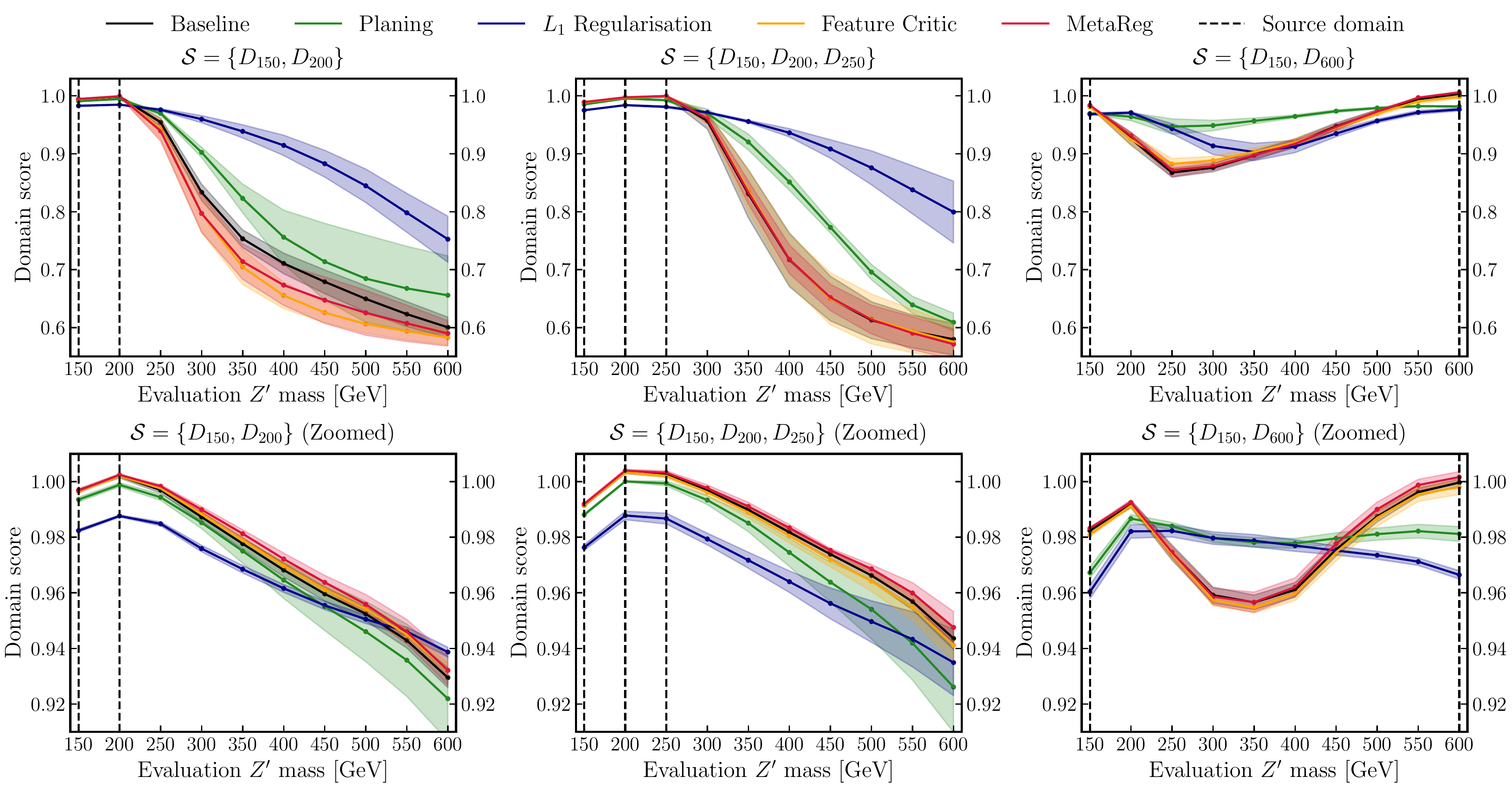}
    \caption{Domain scores for each of the trained models on various source/target
        splits. Source domains are indicated by vertical dashed lines. The top row
        shows results without zooming and the bottom row shows results with zooming 
        (note the change in vertical scale between rows). Curves are the mean of 5
        independent networks with bands displaying one standard deviation around 
        the mean.}
    \label{fig:generalisation-plots}
\end{figure*}

From the right panel we see that each of the performance curves is much flatter
as the domain varies. The inclusion of zooming in the preprocessing steps
significantly improves the generalisation performance of all models. This
indicates that the difference in angular scale between jets of different masses
is a large hindrance to the neural networks' capability to generalise. However,
it remains the case that each model provides poorer classification when
evaluated away from the source dataset. This is not surprising; variation in
the scale of the jets is only a component of the domain shift, and other
factors such as different charged particle multiplicity (for instance) will
also matter.

Next, we present results for models trained on multiple source domains.
Fig.\,\ref{fig:generalisation-plots} contains plots of the models' performance
in the test split of all datasets. Each curve is the mean of the AUC scores of
five models normalised domain-wise by the AUC of a baseline PFN trained only on
the corresponding dataset (the peaks in Fig.\,\ref{fig:baseline-single}).  We
call this metric the \textit{domain score} and in these terms, a perfectly
general model produces a flat line at 1.\footnote{Note that it is possible for
    the domain score to exceed a value of 1.0 since aggregated datasets contain
    more jets.} The bands around the curves show one standard deviation around the
mean. The top and bottom rows of the figure are results without and with
zooming respectively, and the columns correspond to different source masses,
which are shown in each panel by the vertical dashed lines.

Focusing on the top row, where jets are not zoomed, we see that both
meta-learning algorithms exhibit near-baseline performance for
$\mathcal{S}=\{D_{150},D_{200}\}$ and
$\mathcal{S}=\{D_{150},D_{200},D_{250}\}$, while the $L_1$-regularised model
achieves vastly improved generalisation at the cost of slightly poorer
classification on the source masses. The mass-planed network performs somewhat
better than the baseline in these cases, but is the best at interpolating
between distant source domains as in $\mathcal{S}=\{D_{150},D_{600}\}$. The $L_1$ regulariser also bridges the gap better than the
baseline.

The results are different when zooming is applied to the jets. In this case,
all models including the baseline maintain domain scores above 0.9 at all
masses (note the change in scale of the vertical axis compared to the top row).
For $S=\{D_{150},D_{200}\}$ in the bottom-left panel, the meta-learning algorithms and the baseline
slightly outperform the other classifiers, with the $L_1$ regularized model
being the worst classifier at low masses. At high masses all the classifiers
achieve domain scores which are equivalent within the uncertainty bands,
although planing generalises least well.

When another source dataset is added, as in
$\mathcal{S}=\{D_{150},D_{200},D_{250}\}$ in the bottom centre panel, the
results are similar. Finally, when the source domains are chosen at the
extremities so that $\mathcal{S}=\{D_{150},D_{600}\}$, mass-planing exhibits
the most uniform performance across all the domains. However, the meta-learning algorithms and the
baseline have higher domain scores in the vicinity of the source domains but
generalise less well.

\begin{figure*}[tp!]
    \centering
    \includegraphics[width=\textwidth]{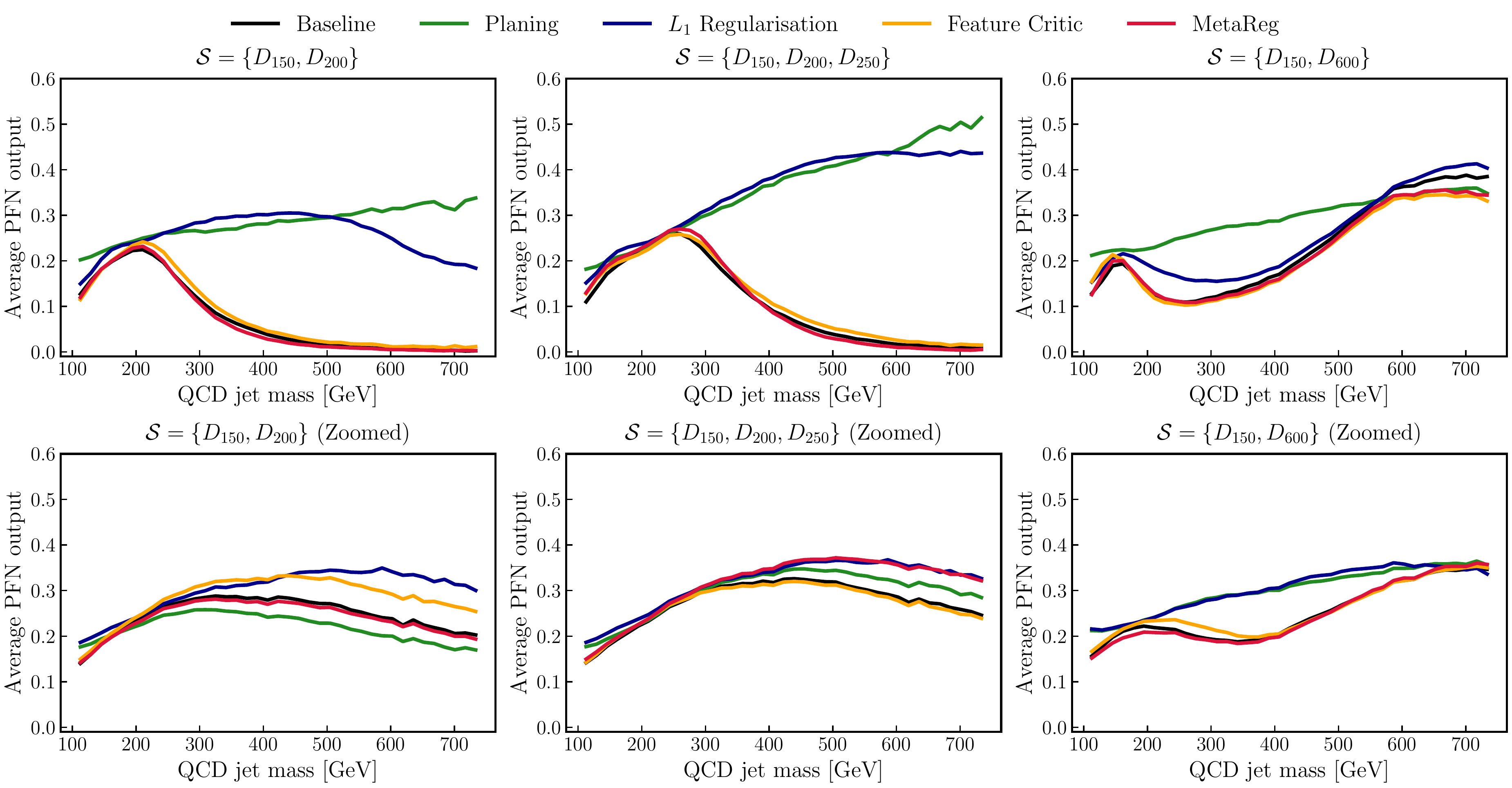}
    \caption{Average PFN output for QCD jets across all masses. Results are shown
        with (bottom row) and without (top row) zooming. The averages are over jets in
        the testing splits of all datasets for 5 independent networks.}
    \label{fig:average-preds}
\end{figure*}

Zooming improves the generalisation of all models. This indicates that the
regularisation and meta-learning methods are unable to completely account for
the variation of jets' angular scales at different masses (or else zooming
would not affect their performance). There is also other information that is
not being fully used by the networks. This is clear from decrease in
performance of the zoomed networks away from the domains where they were
trained. The particle multiplicity in a jet is an example of such information
not captured in the jet angular distribution. Our results show that the
meta-learning models are not optimally leveraging this information relative to
the PFN baseline. While some improvement over the baseline is achieved when
only two source masses are used, the magnitude of this improvement is small
compared to the computational overhead of the meta-learning algorithms.

Given the success of zooming, an alternative approach would be to consider
scale transformations as a symmetry of the data and embed this information into
the network architecture itself. There already exists work on jet taggers that
are equivariant to other symmetries of jets\,\cite{bogatskiy2020lorentz,
    Dolan:2020qkr, Shimmin:2021pkm, Dillon:2021gag} as well as implementations of
scaling-equivariant CNNs\,\cite{DBLP:journals/corr/abs-1910-11093,
    DBLP:journals/corr/abs-1909-11193, Sangalli2021ScaleEN}. Of course, in
applications where the relationship between domains is less easily understood,
it may not be possible to identify the appropriate data augmentation procedure.
Our results suggest that in such a scenario, simple weight regularisation is
able to partially resolve this gap. Alternatively, one can attempt to
\emph{learn} the transformations as in Ref.\,\cite{DIRGAN}.

The fact that MetaReg is often outperformed by standard $L_1$ regularisation
indicates that it is not able to find the optimal regulariser in those cases.
There are a number of potential reasons for this, for example the algorithm may
be particularly sensitive to the hyperparameters that we matched to the
baseline (the learning rate $\alpha$ and the batch size). It may be possible to
improve the performance of MetaReg with a more thorough grid search that includes
the full space of hyperparameters, including the MLP regulariser architecture.
However, such a process is computationally expensive and is not guaranteed to
improve on the far simpler $L_1$ regularisation.

\section{Correlation with jet mass}
\label{sec:correlations}

The performance of mass-planing in Fig.\,\ref{fig:generalisation-plots}
demonstrates that mass-generalisation is not necessarily a byproduct of
decorrelation. In this section, we pose the reverse question: do generalised
models automatically exhibit less correlation with the jet mass than the
baseline? To answer this question, we plot the average prediction score output
by the different PFNs when evaluated on QCD jets across all datasets. In such a
way, networks that are correlated with the jet mass can be identified by the
presence of a peak in the region of the source masses. This peak is what leads
to a background sculpting effect which makes the estimation of systematic
uncertainties difficult, since a fixed selection threshold that intersects the
peak will prefer jets within the peak. In contrast, a network that is not
correlated with mass will produce a flat curve.

Fig.\,\ref{fig:average-preds} presents such plots for different source masses
with and without zooming the jets. The best decorrelation is achieved by the
mass-planed PFN as expected. When jets are not zoomed, the two meta-learning
algorithms behave essentially the same as the baseline, exhibiting a strong
peak at the source masses. However, the $L_1$-regularized PFN is noticeably
decorrelated compared to the baseline. This is interesting given that there is
no special treatment of the jet mass in this approach, only a generic penalty
for large weights in the network.

When jets are zoomed all models behave similarly, with far less correlation
than the unzoomed baseline. In this case, we only observe a difference between
the baseline and mass-planing for $\mathcal{S}=\{D_{150}, D_{600}\}$. In
combination with the results of the previous section, this suggests that
decorrelation from the jet mass is indeed delivered as a byproduct of effective
generalisation. Specifically, the $L_1$-regularised PFN is the most general
model on $\mathcal{S} = \{D_{150}, D_{200}\}$ and $\mathcal{S} = \{D_{150},
    D_{200}, D_{250}\}$ where it is also the least correlated. Similarly, zooming
provides strong generalisation for all models and also leads to relatively
small dependence on jet mass.

\section{Conclusions}
\label{sec:conclusions}

Jet-tagging has become an important area for the application of
machine-learning methods at the Large Hadron Collider. Studies have often been
carried out in narrow ranges of transverse momentum or invariant mass, raising
the question of what is the optimum way to apply these methods over the full
kinematic range available.

In the context of boosted boson tagging at a wide range of masses, we have
studied the interplay between regularisation, mass decorrelation,
preprocessing, and domain generalisation algorithms. Among the existing
techniques we used were $L_1$ regularisation, planing and zooming, where the
zooming procedure was varied to be independent of the mass, $p_T$ or clustering
radius of the jet. We also studied two domain generalisation algorithms based
on meta-learning, namely the MetaReg and Feature-Critic algorithms.

We found that zooming alone is enough to yield strong generalisation. The
meta-learning algorithms only lead to improvement over the baseline when used
in combination with zooming. However, this improvement was minor and limited to
settings with minimal training masses. In the absence of zooming, $L_1$
regularisation lead to the best generalisation. Mass-decorrelation via planing
was most effective at generalising between distant training points.

We have also investigated the correlation with the jet mass of the model
predictions trained under each method. When jets were not zoomed, $L_1$
regularisation led to similar or improved decorrelation from the jet mass
compared to planing despite being ignorant to jet masses. When zooming was
included in the preprocessing steps, all models exhibited relatively little
correlation with jet mass.

While the MetaReg and Feature-Critic displayed limited utility, one could also
study the efficacy of other learning algorithms such as
Refs.\,\cite{masf,dadg,metavib} based on meta-learning, domain-invariant
variational auto-encoders\,\cite{Ilse2020DIVADI}, representation
learning\,\cite{DIRGAN} or scale-equivariant
networks\,\cite{DBLP:journals/corr/abs-1910-11093,
    DBLP:journals/corr/abs-1909-11193, Sangalli2021ScaleEN}. There is also recent
work on `stable' learning which takes a different approach to generalising to
unseen domains \cite{DBLP:journals/corr/abs-2104-07876,
    DBLP:journals/corr/abs-1911-12580}.

The techniques explored here are also applicable to alternate scenarios wherein
a different property of the signal varies across a range. An example case is
resonant/non-resonant dijet classification which could be generalised across
$m_{JJ}$. It may also be possible to explore discrete domains such as the
choice of Monte Carlo event generator, where generalisation would correspond to
learning only generator-non-specific information, with applications to
experimental data, motivated by Ref.\,\cite{shower-uncertainties}. One could
also consider the possibility of unsupervised domain-generalisation, for
instance via the transfer-learning method in Ref.\,\cite{Fan2021ImportanceWA}
where no label information is present in the source domain.

\section*{Acknowledgements}

This work was supported in part by  the Australian Research Council and the
Australian Government Research Training Program Scholarship initiative.
Computing resources were provided by the LIEF HPC-GPGPU Facility hosted at the
University of Melbourne. This Facility was established with the assistance of
LIEF Grant LE170100200.

%


\end{document}